\documentclass[aps,pre,12pt]{article}
\usepackage{graphicx}

\begin{document}
\title{Modes of speciation in heterogeneous space}
\author{Martin Rost$^1$ and Michael L\"assig$^2$\\
$^1${\it \small Abteilung Theoretische Biologie, Universit\"at Bonn, Kirschallee 1,
  53115 Bonn}\\
$^2${\it \small Institut f\"ur Theoretische Physik, Universit\"at zu K\"oln,
  Z\"ulpicher Stra\ss{}e 77, 50937 K\"oln}}
\maketitle
\begin{abstract}
Modes of speciation have been the subject of a century's debate.
Traditionally, most speciations are believed to be caused by 
spatial separation of populations ({\em allopatry}). Recent
observations~\cite{Meyer_1990,Schliewen_1994,Schliewen_2001,Rico_2002}
and
models~\cite{MaynardSmith_1966,Antonovics_1971,Dickinson_1973,Rosenzweig_1978,Turner_1995,Noest_1997,Geritz_1998,Kondrashov_1999,Dieckmann_1999,Doebeli_2000,Slatkin_1980},
show that speciation can also take place in {\em sympatry}.
We discuss a comprehensive model of coupled differentiation in
phenotype, mating, and space, showing that spatial segregation can be
an induced process following a sympatric differentiation. This is
found to be a generic  mechanism of adaptation to heterogeneous 
environments, for which we propose the term {\em diapatric}
speciation \cite{Greek}. It explains the ubiquitous spatial patching
of newly formed species, despite their sympatric
origin~\cite{Schliewen_1994,Schliewen_2001,Rico_2002}.
\end{abstract}

Allopatric speciation occurs in populations extending over a sufficient range
in space and time. If subpopulations become spatially isolated, they can
diverge in phenotype by adaptation to different environments as well as by
genetic drift. A similar divergence is possible while the subpopulations
maintain a limited spatial contact
\cite{Dickinson_1973,Mayr_1963,Gavrilets_2000,Bush_1975,Endler_1977},
which is commonly referred to as {\em parapatric} speciation. Pre- or
post-mating incompatibilities can develop subsequently, leading to
reproductive isolation. Neither the primary phenotypic separation nor the
secondary reproductive isolation require disruptive selection. Hence,
allopatric or classical parapatric speciation may well take too much space and
time to account for radiation events and rapid species divergence
\cite{Dieckmann_2002,Hutchinson_1959}.

In recent years, phylogeographic observations have produced convincing
evidence for speciation in sympatry. Reproductive isolation has occurred in
cichlid populations in African lakes over a few thousand generations
\cite{Meyer_1990,Schliewen_1994,Schliewen_2001}. A salmon population is
reported to have separated within only 14 generations
\cite{Hendry_2000}. Sympatric speciation thus appears to occur rapidly even in
small contiguous environments without spatial barriers.
In theoretical models, it is always driven by {\em disruptive selection}. A
phenotypic split can be favored, for example, if individuals of similar
phenotype compete more strongly than distant ones
\cite{Rosenzweig_1978,Geritz_1998}. In a sexually reproducing population,
however, such splits can only happen if the subpopulations become
reproductively isolated so that the birth of hybrids is suppressed
\cite{MaynardSmith_1966,Rosenzweig_1978,Turner_1995,Slatkin_1980}.
Of course, the sympatric scenario cannot explain the
spatial population structure observed in the
phylogeographic studies. Spatial patching of subpopulations
appears to be ubiquitous. For example, the sister species
of cichlids tend to organize themselves into neighboring
regions \cite{Schliewen_2001,Rico_2002,Ruber_1999}. Another
well-documented case are phytophagous insects, which are
found to evolve  mating assortativity together with
specificity to different host plants~\cite{Bush_1989}.

These observations call for a more comprehensive model that
captures the divergence in phenotypic traits, mating, and
space as a cooperative dynamical process. Only recently IBM
simulations in an extended model space with environmental fitness
gradient have been presented \cite{Doebeli_2003} extending previous
studies with complete spatial mixing \cite{Dieckmann_1999,Doebeli_2000}.

The model discussed here addresses parapatric speciation, i.e.,
generic intermediate cases between sympatry and allopatry. 
It affords a detailed analysis of the dynamics, allowing us to
identify different primary speciation mechanisms and 
their conditions of occurence. (A concise discussion of
classification issues and of the appropriate terminology
can be found in Ref.~\cite{Dieckmann_2002}.)
It turns out that the basic evolutionary forces driving speciation
can be captured by a deterministic ``reaction-diffusion'' approach.
We also discuss the role of stochastic effects as they appear in
individual-based models. In this way, we recover the well-known mechanisms 
of allopatric and sympatric speciation. However, there are
many environments with inhomogeneities on smaller scales in
space and time (such as in the examples quoted above),
where spatial variations prevent sympatry and diffusive
migration prevents allopatry. Adaptative evolution then
operates by a new mechanism, for which we propose the term {\em
diapatric speciation}. The population reaches a final state
of efficient spatial patchiness and phenotypic differentation
without hybrids, which is triggered and sustained by
assortative mating. This is in contrast to the traditional
view of parapatric speciation, where assortative
mating takes a merely secondary role in reinforcing an
existing boundary between emerging species
\cite{Mayr_1963,Gavrilets_2000,Endler_1977,Dieckmann_2002}.

\section*{Model}
We consider a population via its {\em  density} in ``internal'' and
``external'' space, $N \equiv N({\bf x}; {\bf r}; t)$. Internal
coordinates ${\bf x} = (x_1, \dots, x_n)$ denote phenotypic quantities,
e.g., body size, beak length, colour. Internal coordinates can be inherited.
This representation is purely phenotypic. A comparison with explicitly
genetic models is given below. 

External coordinates ${\bf r} = (r_1, \dots, r_d)$ lie in the simplest case
in $d$-dimensional Euclidean space. More complicated geometries, e.g., network
structures of habitat patches in fragmented landscapes, are also possible.
In this work, we focus on habitats with a  gradient in quality for different
phenotypes, which induce a spatial dependence of the optimal phenotype
${\bf x}_{\rm opt} ({\bf r})$ and a  population density 
$N({\bf x},{\bf r};t)$ with a {\em joint} dependence on internal
coordinates ${\bf x}$ and external  coordinates ${\bf r}$.

\begin{figure}[!h]
\begin{center}
\includegraphics[width=12cm]{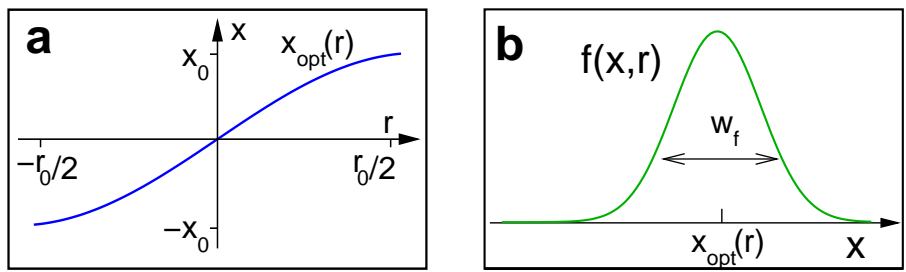}
\end{center}
\caption{\small The fitness landscape of a
  heterogeneous model environment involves a fitness funtion
  $f(x,r)$ that depends on a trait variable $x$ and a
  spatial coordinate $r$. ({\bf a})  The left region
  ($r < 0$) favors smaller values of $x$, the right
  region ($r > 0$) larger ones. The optimal trait
  $x_{\rm opt}(r)$ varies between the values $\pm x_0$ over a
  spatial interval given by the total size $r_0$. ({\bf
  b}) At a given point $r$, the fitness is maximal at 
  $x_{\rm opt}(r)$ and decays rapidly over a
  characteristic scale $w_f$, called the niche width.}
\label{Figfxr}
\end{figure}

In the simplest version of the model we consider one phenotypic coordinate
and a one-dimensional external space of size $r_0$, so $N \equiv N(x,r;t)$
whith $ -r_0/2 \leq r \leq r_0/2$. The phenotype $x$ is directly related
to an ecological fitness or carrying capacity, e.g., with the explicit choice
\begin{equation}
f(x,r) = f(x - x_{\rm opt}(r)) = \exp \left(- \frac{(x - x_{\rm
      opt}(r))^2}{w_f^2} \right)
\end{equation}
which is taken to be constant in time. It decreases with the distance of $x$
from $x_{\rm opt} = x_0 \sin( \pi r/r_0)$, on a scale $w_f$ in phenotype
space. $x_0$ is a measure for habitat heterogeneity and $r_0$ is the spatial
scale of variation. For an illustration see Figure \ref{Figfxr}.

The population $N(x,r;t)$ is subject to the dynamics
\begin{equation}
\partial_t N(x,r;t) = \lambda \partial_r^2 N(x,r;t) + R(x,r;t) + \left( f(x,r)
  - K(x,r;t) \right) N(x,r;t)
\label{EqPopdyn}
\end{equation}
which has the form of a reaction-diffusion equation.

The simplest type of motion in the population is diffusion, in
Eq.~(\ref{EqPopdyn}) appearing as the term $\lambda \partial_r^2 N$, to which
we restrict ourselves in this work. The prefactor defines a length scale in
space, $r_\lambda = \sqrt{\lambda}$, which has to be compared with the habitat
size $r_0$. 

The special case of eq.~(\ref{EqPopdyn}) with $R = 0$ describes the dynamics
of an asexual or clonal population. It similar to the familiar Lotka-Volterra
form. The resource supply $f(x,r)$ and the competition load
\begin{equation}
K(x,r;t) = \int \! dy \; \beta(x,y) \; N(y,r;t),
\end{equation}
which sums up the influence of individuals of trait $y$ on those with
trait $x$, combine to the frequency-dependent {\em fitness} $f - K$.
The competition kernel
\begin{equation}
\beta(x,y) = \beta(|x - y|) = \exp \left( - \frac{|x -
    y|}{w_\beta} \right)
\end{equation}
is maximal for $x \! = \! y$ and decays on a scale $w_\beta$ in internal
space.

Extending this approach to sexually reproducing populations
requires a more detailed model for birth processes, whose
rate itself becomes dependent on the maternal and paternal
population densities. It is convenient to introduce the 
{\em birth excess} per phenotype, space, and time
\begin{equation}
R(x,r;t) = \int \! dy \; dz \; C(x | y,z) \; m(y,z;t) \; N(z,r;t) \; - \;
N(x,r;t).
\end{equation}
by summing over the density of possible mothers $N(z,r;t)$ multiplied by the
probability density $m(y,z)$ for a $z$-female to mate with a $y$-male and the
inheritance probability density $C(x | y,z)$ that this couple will have
offspring of phenotype $x$. The subtracted term $N(r,t)$
describes the total birth rate in the clonal limit. With the
definitions of $C$ and $m$ given below, it is easy to check
that $\int \! dx \; R(x,r) = 0$.
Hence, the excess birth rate describes the net {\em reshuffling} of
population density through sexual reproduction, and $f-K$ remains a
useful measure of the frequency- and space-dependent fitness. The genetic
function $C$ is approximated by a Gaussian, $C(x|y,z) =
\exp(-(x \! - \! \bar x)^2/(2 w_C(\bar x)^2))/\sqrt{2 \pi w_C(\bar x)^2}$,
with $\bar x = (y\! + \!z)/2$, so offspring is distributed near the
mean of the parents' phenotype. Moreover the standard deviation
$w_C(\bar x)$ changes little over the relevant range of phenotypes.
This form can be justified from the hypergeometric model
\cite{Kondrashov_1986,Doebeli_1996,Bulmer_1980,Shpak_1999},
where the quantitative trait $x$ is encoded by $L$ independent two-allele
loci with equal allele frequencies. However, provided the number
of independent loci is sufficiently large, it remains valid more
generally, even if  (i)~the number of loci changes or (ii)~the symmetry between 
the loci is lost~\cite{Barton_2000} because allele frequencies 
change or linkage disequilibria delevop during the speciation process.
Typically this would result in a decrease of $w_C$, but as long as
$w_C < w_f$ and $w_C < w_\beta$, variations in $w_C$ do not influence
the results significantly. See also the discussion at the end of this Section
where we show that this form of $C(x|y,z)$ emerges from a genetically explicit
model quite generically.

Mating preference is crucial for the development of any structure in the
population. Without it, the mating probability is just proportional to the
available males. In this case the entire population is mixing and forms a
single cluster in phenotype, see Figure~\ref{PopDynFixedpoints}(a). This
changes with an affinity of females towards certain types of males,
\begin{equation}
\label{EqMateM}
m(y,z;t) = \frac{\mu(y,z) N(y,r;t)}{\int_w \mu(w,z) N(w,r;t)}.
\end{equation}
Here we restrict ourselves to assortative mate choice by the ecological
trait within a range of width $w_\mu$
\begin{equation}
\mu(y,z) = \mu(|x - y|) = \exp \left( - \frac{|x -
    y|^2}{w_\mu^2} \right).
\end{equation}
With strong enough mating assortativity reproductively isolated subpopulations
can coexist, as shown in Figure~\ref{PopDynFixedpoints}(b).

The population dynamics (\ref{EqPopdyn}) always leads to a stationary density
$\bar N(x,r)$, which reflects the {\em primary selection} given by the fitness
functions $f$ and $K$. On longer, evolutionary time scales, the population
evolves through {\em secondary selection}, i.e., by adaptive mutations
modifying its mating range $w_\mu$~\cite{Karlin_1974}. We study this process
starting from a spatially uniform initial state with random mating. A single
step involves an initially small mutant population that invades the resident
population and eventually becomes a new stationary state $\bar N(x,r)$ with
different trait and mating characteristics. At each step we evaluate whether a
stationary state $\bar N(x,r)$ with given $w_\mu$ is unstable with respect to
a small mutant population $n(x,r;t)$ with different mating range. Successful
mutants are found to invade the resident population completely, producing a
new stationary state. A possible dependence $w_\mu(x)$ due to a linkage
disequilibrium, not taken into account here, is expected only to enhance the
selection pressure towards assortativity.
If adaptive substitutions are sufficiently rare, an evolutionary pathway can
be represented as a sequence of intermediate stationary states leading to an
evolutionary stable final state $\bar N_{\rm es}(x,r)$ \cite{Hammerstein_1996}.
Along the pathway, the number of adaptive steps parametrizes evolutionary
time. More generally,  the mating range $w_\mu$ may be thought of as a further
quantitative trait, the population state being described by a joint
distribution $N(x,w_\mu,r)$. The distribution of $w_\mu$ is strongly peaked,
which justifies the approximation of Eq.~(\ref{EqPopdyn}). The average value
of $w_\mu$ evolves along fitness gradients towards the final state. Generic
evolutionary stable states are found to have either random or strongly
assortative mating. 

Interesting variations in the internal structure of the model are related to
the mating preference. It can depend on
ecologically neutral but inheritable traits such as mating time, marker
traits, and in all cases one may observe phenotypic differentiation
\cite{Kondrashov_1999,Dieckmann_1999,Doebeli_2000,Kriener_2003,Lande_1981}. In
smaller populations some individuals may be unable to mate.
Assortativity restricts the number of possible mates and should be disfavored under
such circumstances. With some modification of Eq.~(\ref{EqMateM}) this effect can
be studied and it turns out that certain types of reproductive isolation are
actually favored \cite{Noest_1997,Kriener_2003}.

Unlike in our model inheritance in sexual population dynamics is often modelled
genetically explicit. The classical approach is to consider a locus with two alleles,
say $a$ and $A$, and under which conditions preferentially homozygous
subpopulations develop \cite{MaynardSmith_1966}. In computer simulations longer
``genomes'' can be used, typically two strings of $L$ bits with the ``alleles''
0 and 1. Genome space is then very large, $2^{2L}$, and a common way to follow the
evolution of a population are simulation of so called individual based models (IBMs),
\cite{Dieckmann_1999,Doebeli_2000,Doebeli_2003}. For their evaluation population
characteristics are sampled over large populations, long times, and many independent runs.

Based on phenotypes but closely related to genetics is the so called hypergeometric
model \cite{Kondrashov_1999,Doebeli_1996,Kondrashov_1986,Shpak_1999}, where
the phenotype of an individual with $2L$ loci is a quantitative trait proportional
to the number of one type of alleles, e.g.\
\begin{equation}
x = \sum_{\nu=1}^{2L} \sigma_\nu \in \{0,1,\dots,2L\},
\label{EqGenoPheno}
\end{equation}
and the alleles are $\sigma_\nu \in \{ 0,1 \}$. If all genotypes mapping onto a
phenotype are equally probable in a population, one can derive the probability
$C(x|yz)$ for a couple with phenotypes $y$ and $z$ to have offspring with $x$:
explicitly for a haploid and to a very good approximation for a diploid genome 
\cite{Doebeli_1996,Kondrashov_1986}. Going one step further away from the
underlying genetic concept leads to models of Quantitative Genetics
\cite{Bulmer_1980} one of which is ours.

Generally such models neglect gene fixation. Also the hypergeometric model
\cite{Kondrashov_1999,Doebeli_1996,Kondrashov_1986,Shpak_1999} may
be invalidated as the central assumption of equiprobability of the various
genotypes contributing to one phenotype can fail \cite{Barton_2000}.
But the same difficulty also arises for IBMs, as e.g.\ in
\cite{Dieckmann_1999,Doebeli_2000,Doebeli_2003}, where only a ``good'' choice
of mutation rate, population and genome size allows for meaningful dynamics
with respect to the question of speciation. It is in these cases, that the
phenotype related hypergeometric model and also quantitative 
phenotypic models as ours behave similarly and thus remain meaningful.

\begin{figure}[!h]
\begin{center}
\includegraphics[width=8cm]{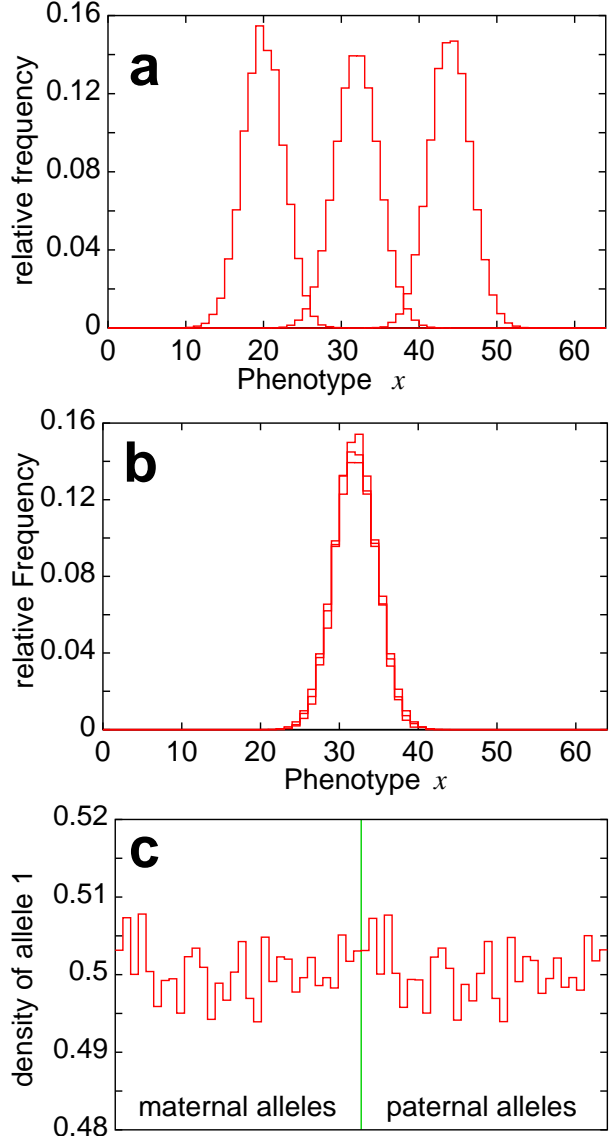}
\end{center}
\caption{\small Examples of $C(x|y,z)$ sampled
  from a simulation of 16384 randomly mating individuals with genome length
  $2L=64$ over $10^7$ generations with mutation probability $10^{-3}$ per
  locus and generation. (a) Examples for $y = z =
  20$, $32$, and $44$. b) Same $z$, but $y = 2L -
  z$. (c) Allele frequencies of 1's at the single loci differ by
  less than 1\% from the average value $1/2$.}
\label{FigIBMHygeo}
\end{figure}

In Figure~\ref{FigIBMHygeo} we show some examples of $C(x|y,z)$ as functions
of the offspring's phenotype $x$ for fixed phenotypes of mother and father,
obtained by sampling over an IBM with genome length $2L = 64$, population size
$16384$, random mating, run for $10^7$ generations. Phenotypes are given by
Eq.~(\ref{EqGenoPheno}). The children's phenotypes are distributed around
values $\bar x(y,z) = (y+z)/2$ with a (nearly Gaussian) distribution whose
widths $w_C$ are practically independent of the parents' phenotypes. In panel
(a) are examples for three values of $z = y$, in panel (b) for values $z =
2L-y$, such that the parents' mean phenotypes are all identical $\bar x = L$.
Panel (c) shows a long time average of the proportion of 1-alleles in the
entire model genome compared to the values observed at the $2L$ model
loci. These simulations show that the elementary combinatorial rules of
inheritance on the genome level used in typical IMB simulations can quite well
be approximated on the phenotype level by continuous functions for
$C(x|y,z)$. The maximum value $\bar x$ and the width $w_C$ may be subject to
corrective terms, but the principle structure of $C(x|y,z)$ remains valid.

In fact it turns out that the precise functional form of the interactions does
not matter too much. We have also studied alternative forms of faster or
weaker decay, e.g., $f(x,r) \sim \exp(-(x/w_f)^4)$, $\beta(x) \sim
\exp(-(x/w_\beta)^2)$, $\mu(x) \sim \exp(-x/w_\mu)$. Important are the length
scales in internal and external space: the inheritance uncertainty $w_C$, the
competition range $w_\beta$, the resource width $w_f$, habitat heterogeneity
$x_0$, extent of habitatvariation $r_0$, and migration range $r_\lambda$.
Their combination and mutual relation decides about reproductive separation of
the population into two or more subpopulations.

\section*{Results}
We first discuss the special case where $f(x,r) \equiv f(x)$ and $N(x,r;t)
\equiv N(x;t)$ do not depend on the spatial coordinate $r$, which requires
$x_0 = 0$, and we consider a spatially  averaged ``mean field'' version of the
model. Also in this limit sympatric speciation can become manifest, analogous
to the results of the IBM in Refs.~\cite{Dieckmann_1999,Doebeli_2000} which is set in a
similar ecological frame. For all evaluations we assumed the ecological
interactions to extend over a wider range than the inheritance uncertainty,
$w_C < w_f$ and $w_C < w_\beta$. During its adaptation the assortativity range
$w_\mu$ varies, but it remains larger than $w_C$. 

\begin{figure}
\begin{center}
\includegraphics[width=12cm]{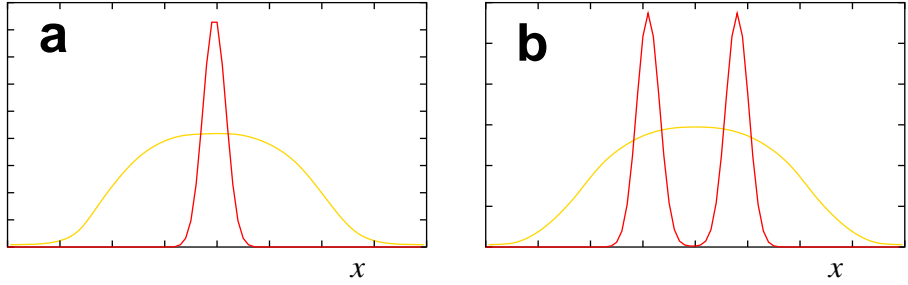}
\end{center}
\caption{\small Fixed point configurations for Equation (\protect
  \ref{EqPopdyn}). (a) weak mating preference gives unimodal population
  structure, (b) strong mating preference allows for bimodal population
  structure. The phenotypic scale is indicated by the resource curve $f(x)$
  plotted in light gray, the vertical scale of population density is
  arbitrary.}
\label{PopDynFixedpoints}
\end{figure}

Evolving reproductive isolation can lead to separation into subpopulations.
Equilibrium profiles of Eq.~(\ref{EqPopdyn}) are shown in
Fig.~\ref{PopDynFixedpoints}, in panel (a) a unimodal population for random
mating, in panel (b) a bimodal under mating assortativity after evolutionary
adaptation of $w_\mu$. For a large enough relative width of the habitat
$w_f/w_\beta > 1.1$ the population evolves into a speciating state as in (b),
otherwise it remains unseparated.

For given parameters $w_C$, $w_\beta$, $w_f$ there may be a range of $w_\mu$,
where both a unimodal (as in Fig.~\ref{PopDynFixedpoints}(a)) and a bimodal
population (as in  Fig.~\ref{PopDynFixedpoints}(b)) are stable fixed points of
Eq.~(\ref{EqPopdyn}). As assortativity gets stronger (decreasing $w_\mu$) the
unimodal profile ( Fig.~\ref{PopDynFixedpoints}(a)) becomes less stable.
It is interesting to note that the transition between a unimodal and a bimodal
population density is not a {\em gradual} process but a {\em fast switch} at a
critical assortativity range $w_\mu$. The switch occurs on the time scale of
population dynamics, much faster than the evolutionary adaptation of
$w_\mu$. Under conditions where the evolutionary stable $w_\mu$ can increase
again (e.g.\ slow variation of $w_f$) one finds hysteresis between the jumps
from uni- to bimodal populations and back.

A population profile as in Fig.~\ref{PopDynFixedpoints}(b) {\em cannot} be a
stable fixed point of the asexual version, i.e., the limit $w_\mu \to 0$, of
Eq.~(\ref{EqPopdyn}): The gap between the two parts of the population would
fill up resulting in a wider unimodal population profile covering most of the
accessible phenotype range \cite{Noest_1997}. When $w_\mu$ remains finite
two peaks in the population profile having widths close to but mutual distance
greater than $w_\mu$ are stabilized by sexual reproduction because it accumulates
offspring closer to their maxima. By this effect sexual reproduction
{\em helps} speciation.

We now turn to the general case. The spatial model dynamics generates
differentiation of the population, which can be measured in two ways: (i) The
{\em mating differentiation index} $\delta$ is defined as the actual rate of
crossmating between two subpopulations at a given point $r$, normalized by the
same rate with random mating. For any population state $N(x,r)$, the mating
differentiation index $\delta$ (at the point $r = 0$ and between the
subpopulations $x<0$ and $x>0$) is defined by 
\begin{equation}
1 - \delta = \frac{1}{Z} \int_{y < 0} \! dy \int_{z > 0} \! dz \;
\frac{1}{2} \left[
m(y,z,0) N(y,0) + m(z,y,0) N(z,0) \right],
\label{defdelta}
\end{equation}
where $Z$ is the same integral evaluated with random mating, i.e., with
$\mu(y,z) = 1$ for all $y,z$. (Analogous measures can be defined for different
$r$ and different trait subpopulations). Here we monitor the two subpopulations
$x < 0$ and $x > 0$ at the boundary between the left and right regions ($r=0$).
(ii) The {\em spatial differentiation index} $\chi$ is defined in terms of the
``trait overlap" between the populations at two different points in space. Phenotypes
$x$ that are intrinsically viable at one of these points are distinguished from
those that are merely advected by diffusion. The spatial differentiation index
$\chi$ (evaluated at the points $-r_0/2$ and $r_0/2$) is defined by
\begin{equation}
1 - \chi = \frac{1}{\tilde Z(r_0/2)} \frac{1}{\tilde Z(-r_0/2)}
\int \! dx \; N_v (x,-r_0/2) N_v(x,r_0/2)
\label{defchi}
\end{equation}
where $\tilde Z(r) = \int \! dx \; N_v (x,r)$. A phenotype $x$ is counted as
intrinsically viable at the point $r$ if $w_C \beta(0) N(x,r)^2 - \lambda
\partial^2 N(x,r)/\partial r^2 > 0$, i.e., if a small nonzero population in
the interval $[x - w_C(x)/2, x + w_C(x)/2]$ could exist even without diffusive
advection. In this case, we set $N_v(x,r) = N(x,r)$, otherwise $N_v(x,r) =
0$. Here we take the points $r = -r_0/2$ in the left region and $r = r_0/2$ in
the right region. Both indices vary between $0$ (no differentiation) and $1$ 
(complete separation).

Following the differentiation in phenotype and space over evolutionary times,
three main mechanisms can be identified. They are distinguished by the
structure of their evolutionary stable final populations $\bar N_{\rm
  es}(x,r)$, measured, for example, by the resulting differentiation indices
$\delta_{\rm es}$ and $\chi_{\rm es}$.
\begin{figure}[!h]
\begin{center}
\includegraphics[width=6cm]{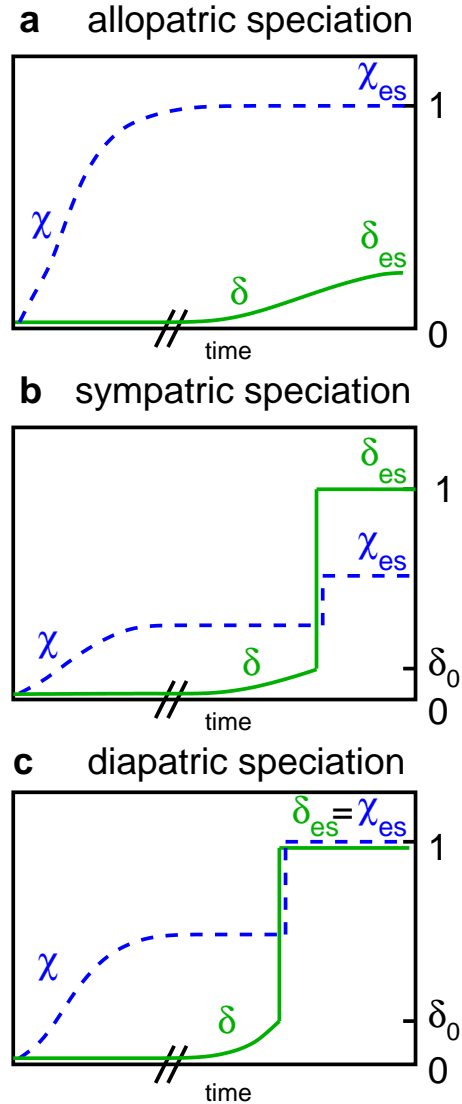}
\end{center}
\caption{\small Three mechanisms of speciation
  can be distinguished by the time dependence
  of the mating differentiation index $\delta$
  and the spatial differentiation index $\chi$ 
  (see text). The initial population has random mating ($\delta = 
  0$) and is spatially homogeneous ($\chi = 0$). 
  Primary selection with random mating (left part of the
  diagrams) is followed by secondary selection on the mating
  range $w_m$ (right part of the diagrams).
  ({\bf a})
  {\bf Allopatric speciation:} Continuous evolution by primary
  selection towards spatial
  segregation ($\chi_{\rm es} = 1$) without reproductive
  isolation ($\delta_{\rm es} < 1$). ({\bf b}) {\bf
  Sympatric speciation:} Discontinuous evolution towards
  reproductive isolation ($\delta_{\rm es} = 1$) without
  spatial segregation ($\chi_{\rm es} < 1$). ({\bf c}) {\bf
  Diapatric speciation:} Discontinuous, cooperative evolution
  towards reproductive isolation ($\delta_{\rm es} = 1$)
  and spatial segregation ($\chi_{\rm es} = 1$).}
\label{FigDefInd}
\end{figure}

\begin{figure}[!h]
\begin{center}
\includegraphics[width=14cm]{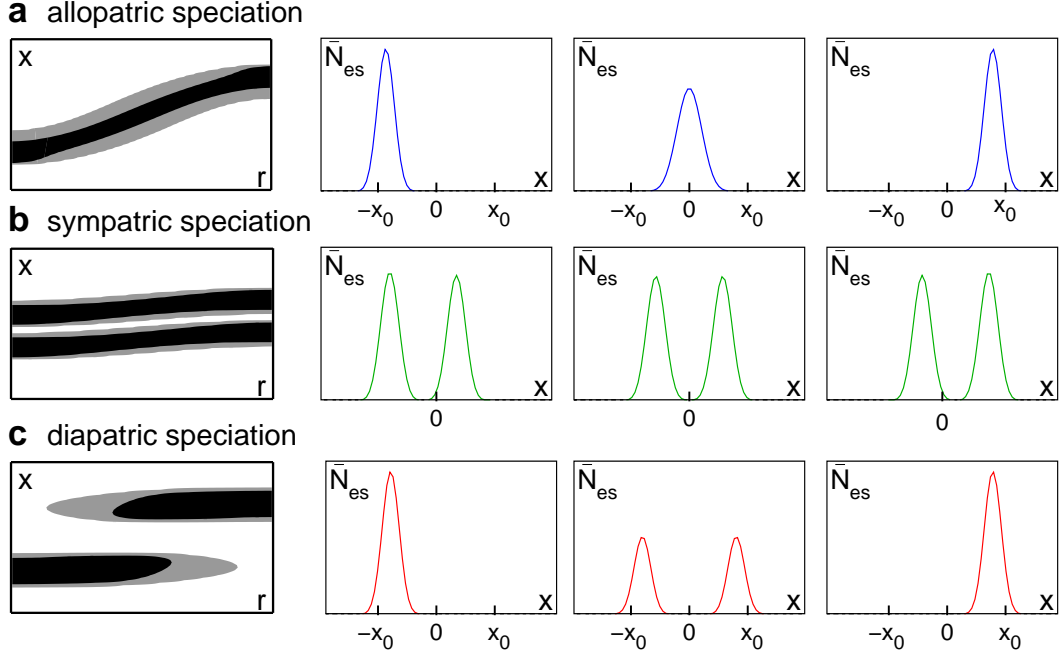}
\end{center}
\caption{\small Evolutionary stable populations after speciation. Left
  column: Populated regions in the $(x,r)$ plane, given by $\bar
  N_{\rm es}(x,r) > 0$. Intrinsically viable phenotypes (shown in black)
  are distinguished from those advected by diffusion (grey). Right
  three columns: trait distributions $\bar N_{\rm es}(x,r\!=\!-r_0/2)$
  (left region), $\bar N_{\rm es}(x,r\!=\!0)$ (boundary between left and
  right region), and $\bar N_{\rm es}(x,r\!=\! r_0/2)$ (right 
  region). ({\bf a}) {\bf Allopatric speciation:} One
  contiguous population cluster, unimodal trait
  distributions, species boundary with hybrids. ({\bf b})
  {\bf Sympatric speciation:} Two disjoint clusters,
  bimodal trait distributions. ({\bf c}) {\bf Diapatric
  speciation:} Two disjoint clusters, trait distributions
  unimodal within the regions and bimodal at the boundary,
  species boundary without hybrids.}
\label{FigPopStruc}
\end{figure}

\begin{figure}[!h]
\begin{center}
\includegraphics[width=12cm]{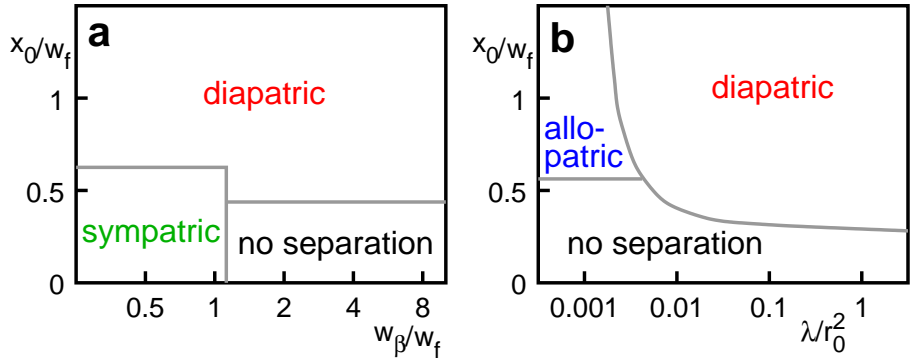}
\end{center}
\caption{\small Phase diagram of speciation,
  specifying the mechanism as a
  function of the effective environment heterogeneity
  $x_0/w_f$, the effective competition range $w_\beta/w_f$,
  and the diffusion coupling between the regions,
  $\lambda/r_0^2$. ({\bf a})~Cross-section in the variables
  $w_\beta/w_f$ and $x_0/w_f$ at fixed $\lambda/r_0^2 =
  0.01$. ({\bf b})~Cross-section in the variables
  $\lambda/r_0^2$ and $x_0/w_f$ at fixed $w_\beta/w_f =
  2.0$. Diapatric speciation is the generic mechanism in
  heterogeneous environments with diffusive coupling.}
\label{FigPhaseDiag}
\end{figure}

{\em Allopatric speciation} shows a gradual increase of the spatial
differentiation index $\chi$ up to $\chi_{\rm es} = 1$, see
Fig.~\ref{FigDefInd}(a). This expresses the patching of small-$x$ phenotypes
into the left region and large-$x$ phenotypes into the right one. The spatial
adaptation of traits involves primary selection by the fitness function
$f(x,r)$ only and occurs independently of mating behavior. Since there is no
sufficient selection pressure towards assortativity, the  mating
differentiation $\delta_{\rm es}$ remains small. The corresponding
evolutionary stable population $\bar N_{\rm es}(x,r)$ is a contiguous cluster
in the $(x,r)$ plane as shown in Fig.~\ref{FigPopStruc}(a). At a given point
$r$, the trait distribution is unimodal and centered around the local fitness
maximum $x_f(r)$. There is a limited gene flow between the large-$x$ and
small-$x$ subpopulations, which is maintained by the intermediate phenotypes
near the boundary ($r = 0$).

{\em Sympatric speciation} is characterized by an increase of mating
differentiation up to complete reproductive isolation; see
Fig.~\ref{FigDefInd}(b). The index $\delta$ jumps discontinuously from a value
$\delta_0 < 1$ to $\delta_{\rm es} = 1$, implying that stationary population
states with $\delta_0 < \delta < 1$ cannot exist. The speciation is driven by
secondary selection involving the frequency-dependent fitness $K$, just as in
previous models of strict sympatry. Spatial variations are irrelevant, and the
spatial differentiation remains incomplete ($\chi_{\rm es} < 1$). The
evolutionary stable population $\bar N_{\rm es}(x,r)$ shown in
Fig.~\ref{FigPopStruc}(b) consists of two disjoint clusters, corresponding to
a bimodal trait distribution at every $r$. The gene flow between these
subpopulations is suppressed by assortative mating. In particular, there are
no hybrids near the boundary ($r = 0$).

{\em Diapatric speciation} is the co-evolution of assortative mating and
spatial segregation by secondary selection. The indices $\delta$ and $\chi$
jump to $\delta_{\rm es} = \chi_{\rm es} = 1$ simultaneously, leading to an
evolutionary stable state with reproductive isolation and patching into the
left and right region, see Fig.~\ref{FigDefInd}(c). Prior to the jump, the
spatial segregation is prevented by diffusive coupling between the regions. It
becomes possible only once reproductive isolation is established. The
population $\bar N_{\rm es}(x,r)$ of Fig.~\ref{FigPopStruc}(c) has two
disjoint clusters. The trait distribution is unimodal within both regions and
bimodal near the boundary; there are again no hybrids. The suppression of the
gene flow between the clusters is now two-fold, by reproductive isolation and
by spatial separation.

Does a population actually speciate, and if so, by which mechanism? This turns
out to depend largely on only three parameters, the effective environmental
heterogeneity $x_0/w_f$, the effective competition range $w_\beta/w_f$, and
the diffusive coupling between the regions, $\lambda/r_0^2$. (Here we have
chosen $w_f$ and $r_0$ as the basic scales in trait space and real space.) The 
``phase diagram'' of Fig.~\ref{FigPhaseDiag} shows the mechanism of speciation
as a function of these parameters. Allopatric speciation is possible only with
a sufficiently large heterogeneity and a sufficiently small diffusive coupling
(i.e., small $\lambda$ or large region size $r_0$). Conversely, sympatric
speciation requires a sufficiently small heterogeneity, as well as a
sufficiently small competition range ($w_\beta/w_f < 1$). Diapatric speciation
involves no restriction on the competition range, that is, it works for
frequency-dependent as well as for density-dependent selection. It is seen to
be the generic mechanism in many realistic environments with heterogeneities
and diffusion.

This compares to the recent results of Ref.~\cite{Doebeli_2003} where
adaptive speciation is seen to generate a sharp geographical
differentiation in an individual based model. Working over a wider
range of parameters the present model is able to relate this diapatric
mechanism to other modes of speciation by identifying their respective
regions in terms of the relevant parameters.

The present model thus allows for a clear identification of the
evolutionary mechanisms underlying speciation, of the dynamical
patterns, and of the resulting population structures. Examples are the
separation indices $\delta$ and $\chi$ (Eqs.~(\ref{defdelta}) and
(\ref{defchi})) and the distinction of intrinsically viable
populations from populations merely advected by diffusion. Of course,
this kind of differential analysis is very difficult in
individual-based models, which always suffer from small discrete
population sizes. On the other hand, the effect of demographic
stochasticity and other fluctuations can also be studied within the
framework of Eq.~(\ref{defdelta}) by adding a stochastic noise
term. The evolutionary stable population densities $\bar N_{\rm
es}(x,r)$ are found to be stable under such perturbations. Stability
or instability of stationary states $\bar N(x,r)$ become immediately
apparent in differential equations such as (\ref{defdelta}) by their
rates of convergence or divergence. The fast transition between
coherent and segregated population states thus explains itself
naturally from a simple analysis.

\section*{Discussion}
In summary, our model suggests that speciation is a
highly cooperative process involving the adaptive
differentiation of a population in its ecological
characters, its mating behavior, and its spatial structure.
Diapatric speciation is the generic mechanism of fully coupled
differentiation. It reduces to allopatric or sympatric
speciation in special cases. All three mechanisms are part
of a unified dynamical picture, conceptually different
from the old dichotomy between sympatry and allopatry.

In classical observations, the spatial separation of newly
formed species has often been regarded as the primary
driving force of the speciation process. Our results call for
a fresh look at the data and may offer a different
interpretation in some cases. The diapatric mechanism involves
spatial separation as an induced process, triggered by the
reproductive isolation of subpopulations. This two-fold
separation in phenotype and space between the emerging
species cuts the gene flow more efficiently than the other
mechanisms, which involve only one kind of separation.

Diapatric species boundaries are established and maintained
by natural selection so no external barriers have to be
postulated. They follow regional boundaries and are
distinguished from the allopatric case by the efficient
suppression of hybrids in the boundary zone during the
primary speciation process. (Secondary reproductive isolation
can suppress hybrids also in allopatry.)
It is crucial to note that reproductively
decoupled populations can adapt to spatial heterogeneities
of smaller size than interbreeding ones. Diapatric
speciation is also remarkably fast, since the loss of
interbreeding takes place through an abrupt change of the
stationary population state as discussed above. This
transition is driven by natural selection, unlike the
secondary mating differentiation mechanisms in allopatry,
which are expected to operate by genetic drift and hence to
be slower. Of course, the genetic fixation of permanent
incompatibilities between the emerging species (postzygotic
isolation) is always slow. Before that point, both
reproductive and the spatial separation are reversible if
the environmental conditions change, as has been confirmed
by recent observations \cite{Seehausen_1997}. Hence,
diapatric splits appear to be an efficient adaptation
mechanism for sexual populations on small scales of space
and time. Most of these splits are wiped out again
on longer time scales, while a few develop into permanent
speciation.

\subsection*{Acknowledgments}
We are grateful to N.\ Barton, A.\ Hastings, M.\ Kirkpatrick, A.\ Kondrashov, and
M.\ Rosenzweig for useful discussions and valuable comments. Particular thanks are
due to D.\ Tautz for numerous comments throughout this work.

Correspondence and requests for materials can be addressed to both authors
(emails: lassig@thp.uni-koeln.de, martin.rost@uni-bonn.de).

\end{document}